\documentclass[twocolumn,showpacs,preprintnumbers,amsmath,amssymb]{revtex4}

\usepackage{graphicx}% Include figure files
\usepackage{dcolumn}% Align table columns on decimal point
\usepackage{bm}% bold math

\begin{document}

\title{On the Potential of the Irreducible Description of Complex Systems \\
for the Modeling of the Global Financial System}

\author{Galina and Victor Korotkikh}
\email{g.korotkikh@cqu.edu.au, v.korotkikh@cqu.edu.au}

\affiliation{School of Computer Science\\
CQUniversity, Mackay\\
Queensland 4740, Australia}

%\date{\today}

\begin{abstract}

The recent financial crisis has sharply revealed that current
understanding of the global financial system is more than limited.
In the recovery plan the confidence in the underlying theory is
crucial. To address the problem we propose the description of
complex systems in terms of self-organization processes of prime
integer relations. The description suggests to consider the global
financial system through the hierarchical network built by the
totality of the self-organization processes. To make the description
operational we propose an integration principle and support it by
computational experiments using a multi-agent system. Remarkably,
based on integers and controlled by arithmetic only the description
raises the possibility to develop an irreducible theory of complex
systems. These may open a fundamentally new perspective for the
modeling of the global financial system.

\end{abstract}


\maketitle

\section{Introduction}

The recent financial crisis tests our approach to reality and
defines us much by the reaction. It has sharply revealed that
current understanding of the financial system, for a rather long
time shaped and entangled by the globalization into a complex
entity, is more than limited. We can not clearly see the way out of
the crisis and, as uncertainty prevails, there is no visible ground
to restore the confidence. At the same time a huge potential of
goodwill has been accumulated recently for a change. In the recovery
plan the confidence in the underlying theory is crucial.

However, since the stakes are extremely high, a certain aspect,
usually not essential at all, has to be made very clear.
Recommendations for the course of actions can be proposed only by
using theoretical considerations derived from basic concepts for
some good reasons assumed to be true. Yet, despite tremendous
progress made in science and technology, so far there is no theory
that originates from irreducible concepts. This actually means that
in principle no theoretical considerations about reality can be
completely trusted, because they are simply consequences of basic
concepts.

In fact the situation is even more demanding. The global financial
system will not be studied in a laboratory, where under different
conditions experiments can be conducted as many times as needed.
Moreover, as a regulatory system of the global economic development
it can actually effect the world as a whole, the trial and error
approach can not be acceptable. It is unlikely much time reserved to
find the solution before it is too late.

\section{Search for an Irreducible Theory of Complex Systems}

A theory, sometimes known as a theory of everything \cite{Barrow_1}
or irreducible \cite{Einstein_1}, and now mainly of interest to
metaphysical debates \cite{Davies_1} and the unification of quantum
mechanics and general relativity \cite{Weinberg_1} may serve the
occasion. However, such a theory appears elusive and highly unlikely
to exist in principle. In fact, as experimental input is always
limited, no matter how successful a theory can be, doubts will
remain.

Yet we believe there is one possibility and it can not be missed:
integers. It is a challenging task to develop a theory that could
operate with statements like $1 + 2 = 3$ and produce results in
agreement with observation. Because such statements are
self-evident, the theory could obtain the irreducible status.

Remarkably, the description of complex systems in terms of
self-organization processes of prime integer relations has been
discovered \cite{Korotkikh_1}-\cite{Korotkikh_4} and showed
promising results \cite{Korotkikh_5}-\cite{Korotkikh_10}. In
particular, it turns out that self-evident integer relations can be
organized into a hierarchical network, where an element of a level
is formed from elements of the lower level so that all elements in
the formation are necessary and sufficient. Moreover, such integer
relations, we call them prime, can encode correlations between
observables and when geometrized can describe the corresponding
dynamics.

The description is realized through the unity of two equivalent
forms, i.e., arithmetical and geometrical. In the arithmetical form
a complex system is characterized by correlation structures built in
accordance with self-organization processes of prime integer
relations. In the geometrical form the processes become
isomorphically represented by transformations of two-dimensional
patterns determining the dynamics of the system. Based on the
integers and controlled by arithmetic only the description raises
the possibility to develop an irreducible theory of complex systems.

Importantly, the role of the description for the development of a
new global financial system could be very special, because its
recommendations would be derived from the concepts we can completely
trust.

\section{On the Role of the Description for the Modeling
of the Global Financial System}

Usually, complex systems are considered in space and time, where
things are assumed to exist separately interacting by forces.
Despite tremendous progress made through this approach, it still has
severe limitations.

First, to understand a complex system in space and time all forces
have to be identified. Only then it is possible to derive the
equation of motion for the system and obtain the needed information.
However, usually not only the precise character of the forces is
unknown, but the forces themselves. This is especially relevant to
the global financial system.

Second, since the forces are not unified in space and time, it is
hard to establish general directions where forces might work
together with certain consequences for complex systems. Without such
directions it becomes difficult to search solutions, as possible
options can not be properly ordered and appear quite similar. In
particular, that may be one of the main reasons why NP-hard problems
become a reality.

Moreover, it is not clear what arguments could be completely trusted
to agree on the objective of the global financial system. Without
this common understanding a long term coordination between different
parties would be problematic.

In its turn our description suggests to consider things through a
new stage rather then in space and time. This stage is the
hierarchical network of prime integer relations - a structure built
by the totality of the processes and existing through the mutual
consistency of its parts. Therefore, in the description things are
viewed as integrated parts of one whole.

Although such a holistic vision is not new \cite{Capra_1},
\cite{Nadeau_1}, yet in the description it becomes unique due to the
irreducible mathematical structure \cite{Korotkikh_2}, which has the
potential to make this perspective operational. Significantly, the
description promises to address the problems of complex systems
mentioned above.

Namely, in our description all forces are managed behind the scene
by a single "force" - arithmetic to serve the special purpose: to
hold the parts of a system together and possibly drive its formation
in one and the same direction to make the system more complex.
Importantly, the forces do not exist separately, but through the
self-organization processes of prime integer relations are unified
and controlled to work coherently in the preservation and formation
of complex systems.

Remarkably, a complex system can be seen as an integrated part
defined by certain processes in the hierarchical network of prime
integer relations, as its spacetime dynamics is given by the
geometry of the processes.

\section{Integration Principle of Complex Systems}

The description suggests us to formulate the following integration
principle:
\medskip

{\it The objective of a complex system defined by self-organization
processes of prime integer relations is to fit precisely into the
processes the system is an integrated part of.}
\medskip

The principle appears as a universal objective of a complex system.
In particular, from its perspective the optimization of a complex
system is about fitting and preserving the position of the system in
the corresponding processes.

The geometrical form of the description provides an important
interpretation of the integration principle. In particular, the
position of a system in the processes can be associated with a
certain two-dimensional shape, which the geometrical pattern of the
system has to take to satisfy the integration principle. Therefore,
this suggests that in the realization of the integration principle
it is important to compare the current geometrical pattern of the
system with the one required for the system by the integration
principle. Since the geometrical patterns are two-dimensional the
difference between their areas can be useful to estimate the result.

Furthermore, the character of the processes may give an efficient
way for the realization of the integration principle. As our
processes work to make systems more complex they move, level by
level, in one and the same direction. In the description the area of
the geometrical pattern can be connected with the information about
the system and thus its entropy. Remarkably, the area should
monotonically increase with the complexity level. Indeed, with each
consecutive level parts of a system combine and make the system more
complex and thus more information is required to characterize the
system.

As with each next level $l < k$  the geometrical pattern of a system
could become closer to the geometrical pattern specified by the
integration principle at level $k$, the performance of the system
might increase but up to level $k$, where it attains the global
optimum, and decrease after as with each consecutive level $l  > k$
the geometrical pattern of the system would differ more from the
required.

Therefore, as the area of the geometrical pattern of a system
increases with the complexity level $l$, the performance of the
system might behave as a concave function of the complexity with the
global optimum attained at the level $k$ specified by the
integration principle.

The following experiments \cite{Korotkikh_5}, \cite{Korotkikh_6}
based on a multi-agent system provide the evidence supporting the
integration principle.

\section{Computational Experiments and Optimality Condition}

Traditionally, when the objective of a complex system can be
formulated explicitly, optimization algorithms can be used to solve
the problem. This provides an important context to test the
integration principle by its realization in finding the global
optimum. For this purpose computational experiments have been
conducted by using a multi-agent system.

In particular, an optimization algorithm ${\cal A}$, as a complex
system, of $N$ computational agents minimizing the average distance
in the travelling salesman problem (TSP) is developed
\cite{Korotkikh_5}.

Let agents start in the same city and choose the next city at
random. Then at each step an agent visits the next city by using one
of the two strategies: random or greedy. In the solution of a
problem with $n$ cities the state of the agents at step $j =
1,...,n-1$ can be described by a binary sequence $s_{j} = s_{1j}
...s_{Nj}$, where $s_{ij} = +1$, if agent $i = 1,...,N$ uses the
random strategy and $s_{ij} = -1$, if the agent $i$ uses the greedy
strategy.

The dynamics of the system is realized by the strategies the agents
choose step by step and can be encoded by an $N \times (n-1)$ binary
strategy matrix
$$
S = \{s_{ij}, i = 1,...,N, j = 1,...,n-1\}.
$$
Remarkably, the matrix can specify $N$ stocks of a financial market
instead of $N$ agents in the TSP problem.

The complexity of the algorithm ${\cal A}$ is tried to be changed
monotonically by forcing the system to make the transition from
regular behavior to chaos by period-doubling. To control the system
in this transition a parameter $v,  0 \leq v \leq 1$ is introduced.
It specifies a threshold point dividing the interval of current
distances passed by the agents into two parts, i.e., successful and
unsuccessful. This information is needed for an optimal if-then rule
\cite{Korotkikh_11} each agent uses to choose the next strategy. The
rule relies on the Prouhet-Thue-Morse (PTM) sequence
$$
+1-1-1+1-1+1+1-1 \ . . .
$$
and has the following description:

1. if the last strategy is successful, continue with the same
strategy.

2. if the last strategy is unsuccessful, consult PTM generator which
strategy to use next.

Remarkably, it is found that for each problem $p$ tested from a
class ${\cal P}$ the performance of the algorithm ${\cal A}$ behaves
as a concave function of the control parameter with the global
maximum at a value $v^{*}(p)$. The global maximums $\{v^{*}(p), p
\in {\cal P}\}$ are used to probe whether the complexities of the
algorithm ${\cal A}$ and the problem are related.

By using the strategy matrices
$$
\{S(v^{*}(p)), p \in {\cal P}\}
$$
corresponding to the global maximums $\{v^{*}(p), p \in {\cal P}\}$
we characterize the geometrical pattern of the algorithm ${\cal A}$
and its complexity. In particular, the complexity ${\cal C}(A(p))$
of the algorithm ${\cal A}$ is approximated by the quadratic trace
$$
C({\cal A}(p)) = \frac{1}{N^{2}}tr(V^{2}(v^{*}(p))) =
\frac{1}{N^{2}}\sum_{i=1}^{N} \lambda_{i}^{2}
$$
of the variance-covariance matrix $V(v^{*}(p))$ obtained from the
strategy matrix $S(v^{*}(p))$, where $\lambda_{i}, i = 1,...,N$ are
the eigenvalues of $V(v^{*}(p))$.

The complexity $C(p)$ of the problem $p$ is approximated by the
quadratic trace
$$
C(p) = \frac{1}{n^{2}}tr(M^{2}(p)) = \frac{1}{n^{2}}\sum_{i=1}^{n}
(\lambda_{i}')^{2}
$$
of the normalized distance matrix
$$
M(p) = \{ d_{ij}/d_{max}, i,j = 1,...,n \},
$$
where $\lambda_{i}', i = 1,...,n$ are the eigenvalues of $M(p)$,
$d_{ij}$ is the distance between cities $i$ and $j$ and $d_{max}$ is
the maximum of the distances.

To reveal a possible optimality condition the points with the
coordinates
$$
\{x = C(p), y = C({\cal A}(p)), p \in {\cal P} \}
$$
are considered. The result indicates a linear relationship between
complexities and suggests the following optimality condition of the
algorithm \cite{Korotkikh_6}.
\medskip

{\it If the algorithm ${\cal A}$ demonstrates the optimal
performance for a problem $p$, then the complexity $C({\cal A}(p))$
of the algorithm ${\cal A}$ is in the linear relationship
\begin{equation}
\label{OPCON} C({\cal A}(p)) = 0.67 C(p) + 0.33
\end{equation}
with the complexity $C(p)$ of the problem $p$.}
\smallskip

According to the optimality condition if the optimal performance
takes place, then in terms of the complexity the dynamics of the
algorithm ${\cal A}$ is in a certain relation with the structure of
the distance network.

Significantly, the optimality condition is a practical tool. Indeed,
for a given problem $p$ by using the distance matrix we can
calculate the complexity $C(p)$ of the problem $p$ and from
$(\ref{OPCON})$ find the complexity $C({\cal A}(p))$ of the
algorithm ${\cal A}$. Then to obtain the optimal performance of the
algorithm ${\cal A}$ for the problem $p$ we need only to adjust the
control parameter for the algorithm ${\cal A}$ to work with the
complexity $C({\cal A}(p))$.

Since the geometrical pattern defines the complexity in our
description, the optimality condition may be interpreted as a
realization of the integration principle. In particular, when the
algorithm  ${\cal A}$ shows the optimal performance for a problem
$p$, there are reasons to suggest that the condition $(\ref{OPCON})$
reads: the geometrical pattern of the algorithm fits exactly into
the geometrical pattern of the problem. The constants in
$(\ref{OPCON})$ are to reconcile different units used in measuring
of the geometrical patterns.

Importantly, due to the concavity of the performance of the
algorithm and polynomial computational complexity of its operations,
in the conducted experiments the NP-hardness of the TSP problem
seems to disappear. This raises the possibility that in our
description NP-hard problems can be avoided \cite{Korotkikh_9}.

\section{Conclusions}

In the paper we have suggested to consider the global financial
system through a new stage rather than in space and time. This new
stage is the hierarchical network - the structure built by the
totality of the self-organization processes of prime integer
relations and existing through the mutual consistency of its parts.
As a result, the global financial system appears not as a separate
entity interacting by forces, but as an integrated part of one
whole, where it finds the meaning and purpose.

To make the description operational we have suggested an integration
principle of complex systems and presented computational evidence
supporting it.

The results raise the possibility to develop an irreducible theory
that for the first time with full confidence in its foundation could
underly the modeling a new global financial system and open a
fundamentally new perspective to resolve the global financial crisis
once and for all.

\bibliography{apssamp}

\end{document}